\documentclass{article}
\usepackage{amsmath, amssymb, amsfonts}
\title{Horizon pressure from junction conditions for Schwarzschild and Rindler geometries} 
\author{Hristu Culetu, \\Ovidius University, Dept.of Physics and Electronics, \\ Mamaia Avenue 124, 900527 Constanta, Romania, \\e-mail : hculetu@yahoo.com}

\begin{document}
\numberwithin{equation}{section}
\pagenumbering{arabic}
\maketitle
\newcommand{\fv}{\boldsymbol{f}}
\newcommand{\tv}{\boldsymbol{t}}
\newcommand{\gv}{\boldsymbol{g}}
\newcommand{\OV}{\boldsymbol{O}}
\newcommand{\wv}{\boldsymbol{w}}
\newcommand{\WV}{\boldsymbol{W}}
\newcommand{\NV}{\boldsymbol{N}}
\newcommand{\hv}{\boldsymbol{h}}
\newcommand{\yv}{\boldsymbol{y}}
\newcommand{\RE}{\textrm{Re}}
\newcommand{\IM}{\textrm{Im}}
\newcommand{\rot}{\textrm{rot}}
\newcommand{\dv}{\boldsymbol{d}}
\newcommand{\grad}{\textrm{grad}}
\newcommand{\Tr}{\textrm{Tr}}
\newcommand{\ua}{\uparrow}
\newcommand{\da}{\downarrow}
\newcommand{\ct}{\textrm{const}}
\newcommand{\xv}{\boldsymbol{x}}
\newcommand{\mv}{\boldsymbol{m}}
\newcommand{\rv}{\boldsymbol{r}}
\newcommand{\kv}{\boldsymbol{k}}
\newcommand{\VE}{\boldsymbol{V}}
\newcommand{\sv}{\boldsymbol{s}}
\newcommand{\RV}{\boldsymbol{R}}
\newcommand{\pv}{\boldsymbol{p}}
\newcommand{\PV}{\boldsymbol{P}}
\newcommand{\EV}{\boldsymbol{E}}
\newcommand{\DV}{\boldsymbol{D}}
\newcommand{\BV}{\boldsymbol{B}}
\newcommand{\HV}{\boldsymbol{H}}
\newcommand{\MV}{\boldsymbol{M}}
\newcommand{\be}{\begin{equation}}
\newcommand{\ee}{\end{equation}}
\newcommand{\ba}{\begin{eqnarray}}
\newcommand{\ea}{\end{eqnarray}}
\newcommand{\bq}{\begin{eqnarray*}}
\newcommand{\eq}{\end{eqnarray*}}
\newcommand{\pa}{\partial}
\newcommand{\f}{\frac}
\newcommand{\FV}{\boldsymbol{F}}
\newcommand{\ve}{\boldsymbol{v}}
\newcommand{\AV}{\boldsymbol{A}}
\newcommand{\jv}{\boldsymbol{j}}
\newcommand{\LV}{\boldsymbol{L}}
\newcommand{\SV}{\boldsymbol{S}}
\newcommand{\av}{\boldsymbol{a}}
\newcommand{\qv}{\boldsymbol{q}}
\newcommand{\QV}{\boldsymbol{Q}}
\newcommand{\ev}{\boldsymbol{e}}
\newcommand{\uv}{\boldsymbol{u}}
\newcommand{\KV}{\boldsymbol{K}}
\newcommand{\ro}{\boldsymbol{\rho}}
\newcommand{\si}{\boldsymbol{\sigma}}
\newcommand{\thv}{\boldsymbol{\theta}}
\newcommand{\bv}{\boldsymbol{b}}
\newcommand{\JV}{\boldsymbol{J}}
\newcommand{\nv}{\boldsymbol{n}}
\newcommand{\lv}{\boldsymbol{l}}
\newcommand{\om}{\boldsymbol{\omega}}
\newcommand{\Om}{\boldsymbol{\Omega}}
\newcommand{\Piv}{\boldsymbol{\Pi}}
\newcommand{\UV}{\boldsymbol{U}}
\newcommand{\iv}{\boldsymbol{i}}
\newcommand{\nuv}{\boldsymbol{\nu}}
\newcommand{\muv}{\boldsymbol{\mu}}
\newcommand{\lm}{\boldsymbol{\lambda}}
\newcommand{\Lm}{\boldsymbol{\Lambda}}
\newcommand{\opsi}{\overline{\psi}}
\renewcommand{\tan}{\textrm{tg}}
\renewcommand{\cot}{\textrm{ctg}}
\renewcommand{\sinh}{\textrm{sh}}
\renewcommand{\cosh}{\textrm{ch}}
\renewcommand{\tanh}{\textrm{th}}
\renewcommand{\coth}{\textrm{cth}}

\begin{abstract}
 We assumed a stress tensor is necessary on the event horizon of a Schwarzschild black hole or on the Rindler horizon for the Israel matching conditions to be satisfied. We found the surface energy density $\rho_{s}$ is vanishing but the surface pressure $p_{s}$ equals $1/16\pi l$ in both cases, where $l$ is the proper distance from the horizon. The junction relations are applied both for $r =$ const. and $T =$ const. surfaces, with the same results for the surface parameters $\rho_{s}$ and $p_{s}$. We emphasize the nonstatic feature of the spacetimes beyond the corresponding horizons.\\

\end{abstract}

\section{Introduction}
The membrane paradigm treats the horizon of a black hole (BH) as a physical 2-dimensional boundary that evolves in a 3-dimensional space \cite{PT}. Price and Thorne replaced the BH horizon by a surrogate ''stretched horizon'' (SH) made from a 2-dimensional viscous fluid. They extend the membrane formalism (introduced earlier by Damour \cite{TD} and Zdanek \cite{LZ1}), to fully dynamical BHs, giving a kinematical description of the structure and evolution of the horizon (a null 3-surface in a 4-dimensional spacetime). 

Parikh and Wilczek \cite{PW} (see also \cite{PK}) introduced an action formulation for the membrane viewpoint. They showed that the boundary term in the derivation of the Euler-Lagrange equations does not vanish on the SH as it does at the boundary of the spacetime. The action formulation has the advantage of providing a link to thermodynamics and quantum mechanics. They chose the SH as a timelike surface just outside the true horizon. It is actually more fundamental than the event horizon (EH) since measurements at the SH represents real measurements that an external observer could perform and report \cite{PW}.

 Ghosh and Perez \cite{GP} consider that the global notion of EH needs to be revised in the context of Quantum Gravity (QG). Even though classically the Tolman temperature on the horizon diverges, in the quantum theory there is a universally (independent of mass) finite local temperature at the horizon. They showed that at $r = 2m+\epsilon$ ($\epsilon << 2m$), the proper distance from the horizon is $l = \sqrt{2m\epsilon}$. Thus, the regular temperature becomes $1/4\pi l$. However, the closest proper distance $l$ must be set by the full quantum theory of gravity.

 Almost in the same time, Frodden, Ghosh and Perez \cite {FGP} introduced a thermodynamical energy $E = A/8\pi l$ and a local surface gravity $\kappa = 1/l$, where $A$ is the horizon area. Once $l$ is set by the QG, $\kappa$ becomes mass-independent, i.e. the local surface gravity is universal. Bianchi \cite{EB} studied the near-horizon geometry of a non-extremal BH as seen by a stationary observer. He introduced the notion of a quantum Rindler horizon in the framework of Loop Gravity. An observer close to the horizon at $l << 2m$ has an acceleration $a = 1/l$ which depends only on the proper distance from the horizon. From here, he obtained the near-horizon energy $E = Aa/8\pi$, the same expression given above. 

By means of the Euclidean action approach, Lemos and Zaslavskii \cite{LZ2} related the membrane model with the entropy of BHs. Their viewpoint is that the SH, as a timelike boundary, is always more convenient in technical terms than the lightlike boundary of the EH. Being infinitely close to the true horizon, one can always take the limit to a lightlike surface. Recently, Widom, Swain and Srivastava \cite{WSS} computed the surface tension of the horizon and showed that all energy and entropy are confined to the horizon surface, using the notion of a vacuum tension of Planck value.

 We try in this paper to show that a stress tensor is needed on the EH in order to be satisfied the junction conditions. We apply the procedure for the Schwarzschild (KS) black hole and for the Rindler spacetime, showing that there is a jump of the extrinsic curvature when the horizon is crossed. With a perfect fluid stress tensor on the SH (viewed as a physical membrane), we found that its energy density is vanishing and the surface pressure equals $1/16\pi l$. Because the line-element is nonstatic beyond the horizon in each case, we write down the matching  conditions both on a constant $r$ and constant time hypersurfaces. 

Throughout the paper the fundamental constants are set to unity, $G = c = \hbar =1$, unless otherwise specified.

\section{Schwarzschild geometry close to the horizon}
Because of the signature flip at the horizon, the exterior KS line-element becomes time-dependent when the EH is crossed \cite{DV, DLC, HC1, EC}
 \begin{equation}
ds^{2} = -\frac{dT^{2}}{\frac{2m}{T}-1} + \left(\frac{2m}{T}-1\right)dR^{2} + T^{2}d\Omega^{2},~~~T < 2m
\label{2.1}
\end{equation}
where the constant $2m$ has been set by Doran et al. \cite{DLC} from a direct confrontation with the exterior KS solution
  \begin{equation}
ds^{2} = -\left(1 - \frac{2m}{r}\right)dt^{2} + \frac{dr^{2}}{1 - \frac{2m}{r}} + r^{2} d \Omega^{2},~~~r > 2m.
\label{2.2}
\end{equation}
$T$ and $R$ in (2.1) plays the role of the timelike and spacelike coordinates, respectively and $d\Omega^{2}$ stands for the line element on the unit 2-sphere. The interior space (2.1) is viewed by an observer at rest, namely $dr = d\theta = d\phi = 0$. 

As suggested by the previous authors, we avoid to work with null surfaces and replace the EHs $r = 2m$ and $T = 2m$ with the so called ''stretched horizons'' (SHs) and write
  \begin{equation}
ds^{2}_{-} = -\frac{dT^{2}}{\frac{2m}{T - \epsilon}-1} + \left(\frac{2m}{T - \epsilon}-1\right)dR^{2} + T^{2}d\Omega^{2},
\label{2.3}
\end{equation}
and
  \begin{equation}
ds^{2}_{+} = -\left(1 - \frac{2m}{r+\epsilon}\right)dt^{2} + \frac{dr^{2}}{1 - \frac{2m}{r+\epsilon}} + r^{2} d \Omega^{2},
\label{2.4}
\end{equation}
where $r = 2m + \epsilon $ and $T = 2m - \epsilon$ are the locations of the two SHs, both of them being very close to the true horizons. Because $\epsilon << 2m$, the proper distance $\rho$ from the horizon may be computed from \cite{HC2}
  \begin{equation}
d\rho^{2} =  \frac{dr^{2}}{1 - \frac{2m}{r}} \approx \frac{2m}{r-2m}dr^{2},~~~for~~ r > 2m,~~r \approx 2m,
\label{2.5}
\end{equation}
which yields $\rho = 2\sqrt{2m(r-2m)} = 2\sqrt{2m\epsilon} = 2l$, where $l$ has been introduced in \cite{GP, FGP, EB}. We shall use from now on their value $l$ for the proper distance from the horizon. That gives us $\epsilon = l^{2}/2m$, with $l << 2m$ and we also assume that $l >> l_{P}$, where $l_{P}$ is the Planck length.

\section{Junction conditions on $r = const.$ surface}
The next task is to find the extrinsic curvature tensors $K_{ab}$ for the surface $r = const.$ in (2.4) and $R = const.$ in (2.3). We have 
 \begin{equation}
 K_{ab}^{+} = -\nabla_{b}n_{a}^{+},~~~~K_{ab}^{-} = -\nabla_{b}n_{a}^{-}
\label{3.1}
\end{equation}
with the normal vectors given by
  \begin{equation}
  n_{a}^{+} = \left(0, \frac{1}{\sqrt{1 - \frac{2m}{r+\epsilon}}}, 0, 0\right),~~~n_{a}^{-} = \left(0, \sqrt{\frac{2m}{T-\epsilon}-1}, 0, 0\right),
\label{3.2}
\end{equation}
 $n_{a}^{+}n^{a}_{+} = 1,~ n_{a}^{-}n^{a}_{-} = 1$ and $a,b$ run from $0$ to $3$. The only nonzero affine connections we need are given by
   \begin{equation}
 \Gamma_{RR}^{T} = \frac{m}{(T - \epsilon)^{2}}\left(1 - \frac{2m}{T - \epsilon}\right),~~~  \Gamma_{\theta \theta}^{T} = -T \left(1 - \frac{2m}{T - \epsilon}\right)
\label{3.3}
\end{equation}
for the interior metric and 
   \begin{equation}
 \Gamma_{tt}^{r} = \frac{m}{(r+ \epsilon)^{2}}\left(1 - \frac{2m}{r+ \epsilon}\right),~~~  \Gamma_{\theta \theta}^{r} = -r \left(1 - \frac{2m}{r+ \epsilon}\right),
\label{3.4}
\end{equation}
for the exterior one. With $n_{a}^{-}$ from (3.2) we find that $K_{TT}^{-} = K_{\theta \theta}^{-} = K_{\phi \phi}^{-} = 0$, as it should be because the interior metric (2.3) does not depend on the radial coordinate $R$. In contrast, for the exterior geometry one obtains (see also \cite{HC2}) 
   \begin{equation}
 K_{tt}^{+} = -\frac{m}{(r+ \epsilon)^{2}} \sqrt{1 - \frac{2m}{r+ \epsilon}},~~~ K_{\theta \theta}^{+} = r \sqrt{1 - \frac{2m}{r+ \epsilon}},
\label{3.5}
\end{equation}
whence
   \begin{equation}
 K^{+} = h^{ab}K_{ab} = \frac{m}{(r+ \epsilon)^{2} \sqrt{1 - \frac{2m}{r+ \epsilon}}} + \frac{2}{r} \sqrt{1 - \frac{2m}{r+ \epsilon}},
\label{3.6}
\end{equation}
evaluated at $r = 2m.~h_{ab} = g_{ab} - n_{a}n_{b}$ is the induced metric on the surface of constant $r$. 

The jump of the second fundamental form (defined by $[K_{ab}] = K_{ab}^{+} - K_{ab}^{-})$ when the horizon is crossed depends only on (3.5) and (3.6), since $K_{ab}^{-} = 0$. It is clear that we need a nonzero stress tensor on the horizon (the SH in our case \footnote{We have in fact two SHs, at $r = 2m + \epsilon$ and $T = 2m - \epsilon$, but when $\epsilon \rightarrow 0$, they tend to the EH $r = T = 2m$. The thin layer between the two has the thickness $(2m+\epsilon) - (2m-\epsilon) = 2\epsilon$, which becomes null when $\epsilon \rightarrow 0$.}) 

The Lanczos equations give us
   \begin{equation}
 [K h_{ab} - K_{ab}] = 8\pi S_{ab},
\label{3.7}
\end{equation}
with $a,b = 0, 2, 3$ and $S_{ab}$ - the surface energy-momentum tensor, which  we take to have a perfect fluid form
   \begin{equation}
  S_{ab} = \rho_{s} u_{a}u_{b} + p_{s}q_{ab}.
\label{3.8}
\end{equation}
In (3.8), $\rho_{s}$ is the surface energy, $p_{s}$ is the surface pressure, $u_{a} = (\sqrt{1 - \frac{2m}{r+\epsilon}}, 0, 0, 0)$ (defined within the hypersurface) is the normal to $t = const.$ surface and $q_{ab} = h_{ab} + u_{a}u_{b}$ is the induced metric on the level surfaces of constant time surface \cite{KKP}.For the $tt$-components of (3.7) we have
   \begin{equation}
 [K h_{tt} - K_{tt}] = - K_{tt}^{+} + h_{tt}K^{+} =    8\pi S_{tt}.
\label{3.9}
\end{equation}
By means of (3.5) - (3.6) and keeping in mind that $\epsilon << 2m$, we get $S_{tt} = 0$, whence $\rho_{s} = 0$. The same procedure leads to
   \begin{equation}
 - K_{\theta \theta}^{+} + h_{\theta \theta}K^{+} = 8\pi S_{\theta \theta} = m\sqrt{\frac{2m}{\epsilon}}.
\label{3.10}
\end{equation}
 In terms of the proper distance from the horizon $l = \sqrt{2m\epsilon}$, the surface pressure becomes $p_{s} = 1/16\pi l$, in the approximation used. A null surface energy $\rho_{s}$ has also been obtained by Kolekar et al. \cite{KKP} in their study of the entropy of spherically symmetric gravitational shells on the verge of forming a BH. $p_{s}$ diverges, of course, when $\epsilon \rightarrow 0$. To avoid that, we could take $l = l_{P}$. Quantum geometry does not give a better choice. However, we consider QG will finally give the best fit for the value of $l$.
 
 \section{Junction conditions on $T = const.$ surface}
 We stress that the inner geometry (2.1) is time-dependent and one should be useful to write down the junction conditions on a $T =$ const. surface, to check whether the values of $\rho_{s}$ and $p_{s}$ obtained above are preserved. The normal vectors to $T =$ const. hypersurface read as
   \begin{equation}
  n_{a}^{-} = \left( \frac{1}{\sqrt{\frac{2m}{T-\epsilon}-1}},0 , 0, 0\right),~~~n_{a}^{+} = \left( \sqrt{1 - \frac{2m}{r+\epsilon}}, 0, 0, 0\right).
\label{4.1}
\end{equation}
with $n_{a}^{+}n^{a}_{+} = -1,~ n_{a}^{-}n^{a}_{-} = -1$. As in the previous situation, the extrinsic curvature $K_{ab}^{+} = 0$ since the exterior metric is time-independent. We have now $u_{a} = (0, \sqrt{ \frac{2m}{T-\epsilon}-1}, 0, 0)$, defined within the hypersurface $T = const.$ and normal to $R = const.$ surface. In addition, $q_{ab} = h_{ab} - u_{a}u_{b}$.

 For the inner metric we get
    \begin{equation}
 K_{RR}^{-} = -\frac{m}{(T- \epsilon)^{2}} \sqrt{ \frac{2m}{T- \epsilon}-1},~~~ K_{\theta \theta}^{-} = T \sqrt{ \frac{2m}{T- \epsilon}-1} = \frac{K_{\phi \phi}^{-}}{ sin^{2}\theta},
\label{4.2}
\end{equation}
 whence
   \begin{equation}
  K^{-} = h^{ab}K_{ab}^{-} = -\frac{m}{(T- \epsilon)^{2} \sqrt{ \frac{2m}{T- \epsilon}-1}} + \frac{2}{T} \sqrt{ \frac{2m}{T- \epsilon}-1}, 
\label{4.3}
\end{equation} 
 evaluated at $T = 2m$. We have now $h_{ab} = g_{ab} + n_{a}n_{b}$ - the induced metric on the surface $T = const.$. To find $\rho_{s}$ and $p_{s}$ we need
    \begin{equation}
 [- K_{\theta \theta}^{-} + h_{\theta \theta}K^{-}] = \frac{mT^{2}}{(T- \epsilon)^{2} \sqrt{ \frac{2m}{T- \epsilon}-1}} - T \sqrt{ \frac{2m}{T- \epsilon}-1} 
\label{4.4}
\end{equation}
 and
 \begin{equation}
 [- K_{RR}^{-} + h_{RR}K^{-}] =  - \frac{2}{T} (\frac{2m}{T- \epsilon}-1)^{3/2}.
\label{4.5}
\end{equation}
 For $\epsilon << 2m$ and with $T = 2m$, one obtains from (4.5) that $\rho_{s} = 0$, when $\epsilon \rightarrow 0$. In the same approximation, the r.h.s. of (4.4) gives $m\sqrt{2m/\epsilon} - 2m\sqrt{\epsilon/2m}$. With $\epsilon << 2m$ only the first term survives , which leads to $8\pi S_{\theta \theta} = m\sqrt{2m/\epsilon} $. Therefore, $8\pi p_{s} = (1/4m)\sqrt{2m/\epsilon}$, or $p_{s} = 1/16\pi l$, as in the previous case.
 
 A similar relation with (2.5) 
   \begin{equation}
d\tau^{2} =  \frac{dT^{2}}{ \frac{2m}{T}-1} \approx \frac{2m}{2m - T}dT^{2}
\label{4.6}
\end{equation}
 gives us the expression of the proper distance $\tau$ to the horizon $T = 2m$
    \begin{equation}
  \tau = -2\sqrt{2m(2m - T)} = -2\sqrt{2m\epsilon} = -2l,  
\label{4.7}
\end{equation}
 with $\tau \in(-4m,0)$. To summarize, from the point of view of an internal observer, located close to $T = 2m$, the surface stress tensor looks similar with that viewed by an external observer.

\section{Rindler geometry close to the horizon}
We take the static Rindler geometry for an accelerating observer, close to the horizon,  to be given by
   \begin{equation}
ds^{2}_{-} = - [1 - 2g(X-\epsilon)] dT^{2} + \frac{dX^{2}}{1 - 2g(X-\epsilon)} + dy^{2} +  dz^{2}
\label{5.1}
\end{equation}
where $T$ is not related to the previous one and $g$ is the constant Rindler acceleration. To pass over the event horizon $X = 1/2g$ we have to keep in mind that $X$ becomes timelike there and $T$ - spacelike. Hence, the new metric becomes time-dependent
   \begin{equation}
ds^{2}_{+} = - \frac{dt^{2}}{2g(t+\epsilon)-1} + [2g(t+\epsilon)-1]dx^{2} + dy^{2} +  dz^{2},
\label{5.2}
\end{equation}
where $t > (1/2g) - \epsilon$. As in the previous example, we get rid of the EHs at $X = 1/2g$ and  $t = 1/2g$ with the help of a tiny $\epsilon <<1/2g$ and so the EH is shifted to the SH at $X = (1/2g) - \epsilon$ and $t = (1/2g) +\epsilon$, respectively. 

We will make use of the following nonzero Christoffel symbols for the two geometries (5.1) and (5.2)
   \begin{equation}
 \Gamma_{TX}^{T} = - \Gamma_{XX}^{X}  = - \frac{g}{1 - 2g(X-\epsilon)},~~~\Gamma_{TT}^{X} = -g[1 - 2g(X-\epsilon)]
\label{5.3}
\end{equation}
and, respectively,
   \begin{equation}
 \Gamma_{tx}^{x} = - \Gamma_{tt}^{t}  =  \frac{g}{ 2g(t+\epsilon)-1},~~~\Gamma_{xx}^{t} = g[2g(t+\epsilon)-1].
\label{5.4}
\end{equation}
It is worth noting that $ 1/2g$ plays here the role of $2m$ from the KS case, so that $\epsilon = l^{2}/(1/2g) = 2gl^{2}$, with $l << 1/2g$. For example, the position of the SH is, respectively,
 \begin{equation}
X = \frac{1}{2g}[1+(2gl)^{2}],~~~and ~~~t = \frac{1}{2g} [1-(2gl)^{2}]
\label{5.5}
\end{equation}

\section{Junction conditions on $X$ = const. surface}
What we have to do now is to follow the steps from chap.3, adapted to the new situation. The normal vectors corresponding to $X$ = const. surface and, respectively, to $x$ = const. are given by
   \begin{equation}
  n_{a}^{-} = \left( 0, \frac{1}{\sqrt{1 - 2g(x-\epsilon)}},0 , 0,\right),~~~n_{a}^{+} = \left(0, \sqrt{2g(t+\epsilon)-1}, 0, 0,\right).
\label{6.1}
\end{equation}
Noting that the metric (5.2) depends on time only and so $ K_{ab}^{+} = 0$, for any $a,b = 0, 2, 3$. However, in the spacetime (5.2) and with $n_{a}^{-} $ from (6.1) we get
    \begin{equation}
 K_{TT}^{-} = -g \sqrt{1 - 2g(X-\epsilon)},~~~ K_{yy}^{-} =  K_{zz}^{-} = 0
\label{6.2}
\end{equation}
and
 \begin{equation}
 K^{-} = - \frac{g}{ \sqrt{1 - 2g(X-\epsilon)}},
\label{6.3}
\end{equation}
evaluated at $X = 1/2g$. 
We find again that $\rho_{s} = 0$ on the Rindler horizon $X = 1/2g$, assumed to be endowed with a perfect fluid stress tensor, as in Eq. (3.8). We are actually constrained to impose  surface stresses at $X = 1/2g$ in order to justify that  $[K_{yy}^{-}] =  [K_{zz}^{-}] \neq 0$, as we shall see. We have, indeed
   \begin{equation}
 [K h_{yy} - K_{yy}] = -K^{-} = \frac{g}{ \sqrt{1 - 2g(X-\epsilon)}} =  8\pi S_{yy},
\label{6.4}
\end{equation}
taken at $X = 1/2g$. Therefore 
   \begin{equation}
	p_{s} = \frac{1}{8\pi}\sqrt{\frac{g}{2\epsilon}} = 1/16\pi l,
\label{6.5}
\end{equation}
which coincides with the previous results (we remind that an equivalence between $2m$ and $1/2g$, the EHs' positions for the KS and Rindler spacetimes, was assumed).

\section{Junction conditions on $T$ = const. surface}
We have now $ K_{ab}^{-} = 0$ as the line-element (5.1) is static
   \begin{equation}
	K_{ab}^{-} = \frac{1}{2\sqrt{-g_{00}}}\frac{\partial h_{ab}^{-}}{\partial T} = 0.
\label{7.1}
\end{equation}
The normal vectors are now
   \begin{equation}
  n_{a}^{+} = \left( \frac{1}{\sqrt{ 2g(t+\epsilon)}-1},0 ,0, 0,\right),~~~n_{a}^{-} = \left(\sqrt{1-2g(x-\epsilon)}, 0, 0, 0,\right).
\label{7.2}
\end{equation}
With $ n_{a}^{+}$ from (7.2) one obtains
    \begin{equation}
 K_{xx}^{+} = g \sqrt{ 2g(t+\epsilon)-1},~~~ K_{yy}^{+} =  K_{zz}^{+} = 0,~~~ K^{+} = h^{ab}K_{ab}^{+} = \frac{g}{\sqrt{ 2g(t+\epsilon)-1}}.
\label{7.3}
\end{equation}
$u_{a}$ is now given by $u_{a} = (0, \sqrt{ 2g(t+\epsilon)-1}, 0, 0)$. The $xx$-component of the jump of the l.h.s. of (3.7) is vanishing and, therefore, $\rho_{s} = 0$. Nevertheless, the $yy$- and $zz$-components give us
   \begin{equation}
 [K^{+} h_{yy} - K_{yy}^{+}] = K^{+} = \frac{g}{ \sqrt{ 2g(t+\epsilon)-1}},
\label{7.4}
\end{equation}
taken at $t = 1/2g$. Hence, for $\epsilon << 1/2g$,
   \begin{equation}
	p_{s} = \frac{1}{8\pi}\sqrt{\frac{g}{2\epsilon}} = 1/16\pi l,
\label{7.5}
\end{equation}
which is coincident with (6.5). Its order of magnitude depends on what QG will say about the value of $l$.

\section{Conclusions}
We emphasized in this paper it is mandatory to have a nonzero pressure on the KS or Rindler horizons for the junction conditions to be obeyed. That is related to the signature flip on the EH so that the geometry becomes nonstatic beyond it. By means of the SH (located close to the EH, at a proper distance $l$), we found the parameters of the perfect fluid ($p_{s}$ and $\rho_{s}$).The surface energy $\rho_{s}$ was found to be zero but the surface pressure equals $p_{s} = 1/16\pi l$. As Ghosh and Perez \cite{GP} have noticed, classically one has $l \rightarrow 0$ and $\epsilon \rightarrow 0$, but quantum-mechanically the proper distance $l$ must be given by the smallest length scale given by quantum geometry. We did not treat the first junction conditions at the horizon because one simply replaced the timelike with spacelike coordinates and vice versa. In addition, as it was shown in \cite{HC2}, the EH is two-dimensional, $ds^{2}_{H} = 4m^{2} d\Omega^{2}$, and so when  $\epsilon \rightarrow 0$, the other two coordinates become useless. In spite of the different geometrical structures, we notice that $p_{s}$ acquires the same value in KS and Rindler cases. That was possible as we have used the equivalence between KS horizon at $r = 2m$ and Rindler horizon at $x = 1/2g$. This is possible if we remind that, near the BH horizon, the KS geometry becomes Rindler.


\begin{thebibliography} {17} 

\bibitem{PT}
R. N. Price and K. S. Thorne, Phys. Rev. D33, 915 (1986).
\bibitem{TD}
T. Damour, Phys. Rev. D18, 3598 (1978).
\bibitem{LZ1}
L. Znajek, Mon. Not. R. Astron. Soc. 185, 833 (1978).
\bibitem{PW}
M. K. Parikh and F. Wilczek, Phys. Rev. D58, 064011 (1998) (arXiv: gr-qc/9712077).
\bibitem{PK}
M. K. Parikh and J. Khoury, Report CU-TP-1167, IUCAA-55/2006 (arXiv: hep-th/0612117).
\bibitem{GP}
A. Ghosh and A. Perez, arXiv: 1107.1320 [gr-qc]; DOI: 10.1103/PhysRevLett.107.241301.
\bibitem{FGP}
E. Frodden, A. Ghosh and A. Perez, arXiv: 1110.4055 [gr-qc].
\bibitem{EB}
E. Bianchi, arXiv: 1204.5122 [gr-qc].
\bibitem{LZ2}
J. P. S. Lemos and O. Zaslavskii, Phys. Rev. D84, 064017 (2011) (arXiv: 1108.1801 [gr-qc]).
\bibitem {WSS}
A. Widom, J. Swain and Y. N. Srivastava, arXiv: 1602.03057 [gr-qc].
\bibitem{DV}
D.N. Vollick, Gen. Rel. Grav. 35, 1511 (2003); \textit{arXiv}: hep-th/0102187.
\bibitem {DLC}
R. Doran, F.S.N. Lobo and P. Crawford, Found. Phys. 38, 160 (2008) (arXiv: gr-qc/0609042).
\bibitem{HC1}
H. Culetu, Cent. Eur. J. Phys. 6, 317 (2008); \textit{arXiv}: hep-th/0703168.
\bibitem{EC}
A. Edery and B. Constantineau, Class. Quantum Grav. 28, 045003 (2011) (arXiv: 1010.5844 [gr-qc]).
\bibitem{HC2}
H. Culetu, Phys. Lett. A376, 2817 (2012) (arXiv: 1103.2645 [gr-qc]).
\bibitem{KKP}
S. Kolekar, D. Kothawala and T. Padmanabhan, Phys. Rev. D85, 064031 (2012) (arXiv: 1111.0973 [gr-qc]).
\bibitem{HC3}
H. Culetu, Phys. Lett. B703, 641 (2011) (arXiv : 1011.3343 [gr-qc]); arXiv: 1101.2980 [gr-qc].


\end{thebibliography}
\end{document}